\documentclass[journal]{IEEEtran}
\usepackage{amsmath, amssymb, bbm}
\usepackage{amsthm}
\usepackage{graphicx,epsfig}
\usepackage{subfigure}
\usepackage{verbatim}
\usepackage{soul}
\usepackage{color}

\newtheorem*{proof*}{Proof}

\usepackage{hhline}
\usepackage{multirow}
\usepackage{caption}

\usepackage{algorithm,algorithmic}

\usepackage{cite}

\begin{document}

\bstctlcite{IEEEexample:BSTcontrol}

\title{Steady-state Voltage Profile and Long-term Voltage Stability of Electrified Road with Wireless Dynamic Charging}
\author{Chuan Wang, Hung D. Nguyen

\thanks{Chuan Wang and Hung D. Nguyen are with the School of Electronic and Electrical Engineering, Nanyang Technological University, Singapore e-mail: chuan002@e.ntu.edu.sg and hunghtd@ntu.edu.sg.}}

\markboth{ACM’19, Phoenix, AZ, USA, June 2019}%
 {Shell \MakeLowercase{\textit{et al.}}}%

\maketitle

\begin{abstract}
Wireless dynamic charging technologies are becoming a promising alternative solution to plug-in ones as they allow on-the-move charging for electric vehicles. From a power network point of view, this type of charging makes electric vehicles a new type of loads--the moving loads. Such moving loads are different from the traditional loads as they may change their locations constantly in the grids. To study the effect of these moving loads on power distribution grids, this work focuses on the steady-state analysis of electrified roads equipped with wireless dynamic charging. In particular, the voltage profile and the long-term voltage stability of the electrified roads are considered. Unusual shapes of the voltage profile are observed such as the half-leaf veins for a one-way road and the harp-like shape for a two-way road. Voltage swings are also detected while the vehicles move in the two-way road configuration. As for the long-term voltage stability, continuation power flow is used to characterize the maximum length of a road as well as the maximum number of vehicles that the road can accommodate. 
\end{abstract}


\section{Introduction}
Electric vehicles (EVs) becomes an integral part of urban transportation \cite{Hines}. They may overtake conventional vehicles with combustion engines soon owing to having advantages making no noise, releasing no pollution at the vehicle level, and having higher well-to-wheels efficiency. EVs can be further classified as stationary and dynamic charging vehicles according to the charging method. Dynamic charging method allows charging in motions so that it resolves the issue of long charging time in the stationary counterpart. Among dynamic charging technologies, wireless dynamic charging (WDC) is emerging as it can charge vehicles on the move without connecting directly to power cables \cite{economic,electrified,world}.

Currently, the research trends focus on the charging technologies mostly in the device level such as maintaining charging efficiency, regulating battery capacity, and improving economic benefits \cite{multiple-route, ORNL, review}. The effects of dynamic charging in the system level have not been studied extensively. Interestingly, a relevant research area, however, is railway electrification systems with a focus on electric locomotives. Most importantly, \cite{novel,sequential,load,power_flow,railway,optimising} carried out power flow analysis of electrified tracks, wherein several new power flow algorithms are introduced, based on which power supply capacity of traction power supply systems and voltage distribution in transmission systems can be analyzed. Another important research direction on trains, tramways, and buses is to apply regenerative braking energy and leverage the topology of the power supply system to enhance the system's efficiency and economic operation \cite{theimpact, energy}. These works focus on the energy consumption and control methods instead of the impact of moving vehicles on the voltage level and the stability of distribution system. 

While these aforementioned works also consider similar radial network configurations, no existing work studies the steady-state voltage profiles and the long-term voltage stability limits of electrified roads when vehicles are in motion. The traditional loads in power systems can vary the demands but resize at fixed locations in the network diagram. WDC vehicles, on the other hand, may consume the same amount of power, if they travel at a constant speed, but change the positions over time. The advent of these new moving loads may alter the distribution system operation and thus calls for new studies on this topic. This paper, therefore, will construct a series of new patterns about the voltage profile along the electrified roads over time to study the voltage level change while the vehicles travel. This pattern can be used in voltage compensation design in order to maintain voltage levels across the network within a range specified by voltage regulation standards. Another important aspect of steady-state analysis considered in this work is the long-term voltage stability (LTVS) of the electrified road, which concerns the existence of a steady-state solution \cite{HungTuritsyn2015Multistability,nguyen2018framework, EssieHungKostya,dvijotham2017solvability,molzahn2013sufficient,aolaritei2017distributed,dinh2013new}. While typical LTVS limits in power systems are regarded as the maximum loading level that the power line can support, the LTVS thresholds here can be quantified in terms of the maximum road length and the number of vehicles ``safely'' allowed to operate on the electrified roads equipped with WDC. We introduce a novel experimental setup below, relying on a widely-used power software package in the community, i.e., MATPOWER \cite{matpower}, to perform our analysis.

\section{WDC EVs and Power Flow Study} \label{sec:pf}
\subsection{Wireless Dynamic Charging System}
This paper considers a simple dynamic charging EV system. The power source, which is located at the beginning of the electrified road, can be a generator or a coupling point connecting to a distribution grid. Then the power cable line under the electrified road connects power tracks, which are typically configured in series, to the power source. While a vehicle is moving over a power track, its battery can be recharged via the receiving resonator.

\begin{figure}[t]
    \centering
    \includegraphics[width=1\columnwidth]{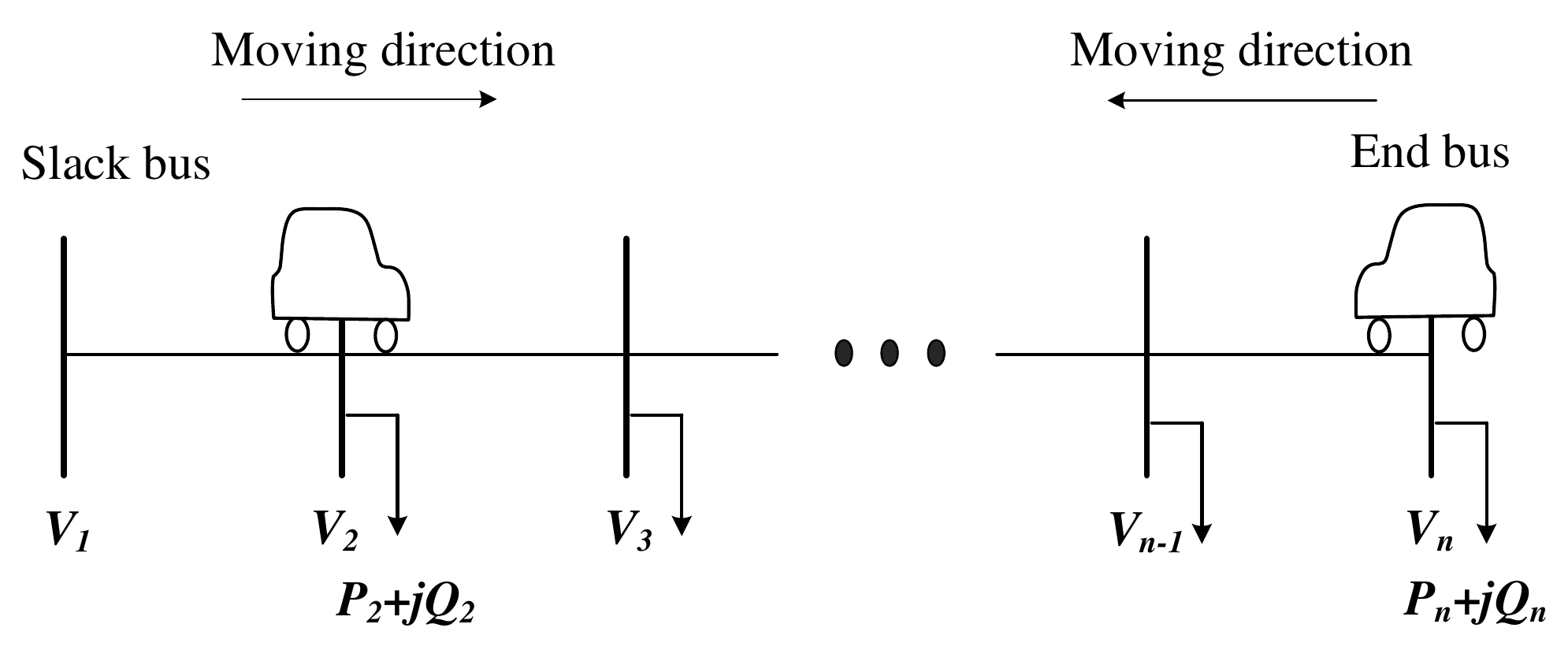}
	\caption{The diagram of simplified WDC system}
	\label{fig:simWDC}
\end{figure}

For simplicity, we assume that all EVs are homogeneous, and they travel at a constant speed with a fixed amount of power. While standard power system packages, such as MATPOWER \cite{matpower}, do not handle moving constant-power (PQ) buses, we propose to split the electrified road into equal-length segments marked with fictitious nodes such that, in each ``snapshot'' of power flow study at a time step, the fleet of vehicles arrives at a new set of fictitious nodes. Such a new set of fictitious nodes is then modeled as PQ buses whereas, except the reference bus, other nodes without EVs consume no power. As the EVs travel along the road, the set of PQ buses also changes. To understand how on-the-move vehicles affect the voltage profile along the road, we repetitively solve power flow problem for such a series of time moments characterized by the set of PQ buses. Then the WDC system can be modeled, at one time-step, as a normal radial grid shown in Figure \ref{fig:simWDC} and can be studied using power system packages. For the steady-state analysis, we find the network voltage solution by solving power flow equations that are presented in Appendix \ref{sec:apppf}.
\section{Voltage profile}\label{sec:volpro}

The parameters of EVs and the WDC system are as following: active power per vehicle ($P_{EV}$): 30 kW; reactive power per vehicle ($Q_{EV}$): 15 kVAr; resistance of cable line: $0.568 \Omega/$km; reactance of cable line: $0.133\,\Omega/$km; on-board PV panel: 2 kW; on-board capacitor: 100 kVAr; and fixed capacitor bank: 300 kVAr. As no common standard of EVs or WDC exists, we rely on several existing works in the literature that present the parameters of the used prototypes \cite{impact,simplified,range,capacitor}. For the cable, we use typical 6-kV distribution cable presented in \cite{cable}. In section \ref{sec:volpro}, the road length and cable line are set to be 2 km. Ten fictitious nodes are evenly distributed on the road. The voltage profiles along this electrified road are presented in the following sections where we consider both one-way and two-way roads. Each moving direction has only one lane. 
\subsection{One-way Road}
In this case, a fleet of two EVs moves from the slack bus to the end bus. At the first time step, the two EVs start from node $1$ and $2$.
\subsubsection{DWC EVs system without voltage compensators}
\begin{figure*}
\centering
\begin{minipage}[b]{.45\textwidth}
    \includegraphics[width=1\columnwidth]{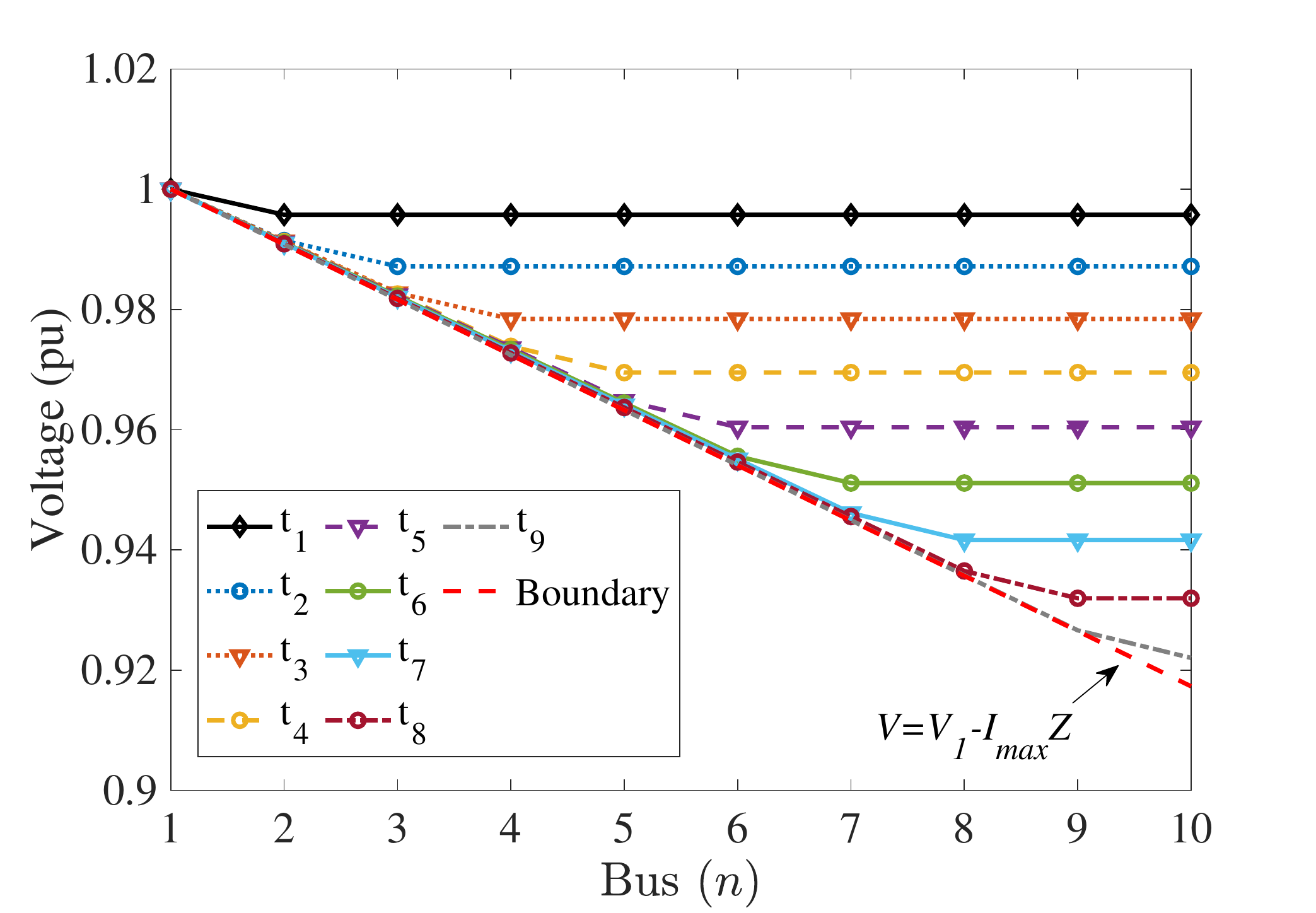}
    \caption{Voltage profile of WDC system along the electrified road (one direction)}
    \label{fig:voltage_one_direction}
\end{minipage}\qquad\qquad
\begin{minipage}[b]{.45\textwidth}
    \includegraphics[width=1\columnwidth]{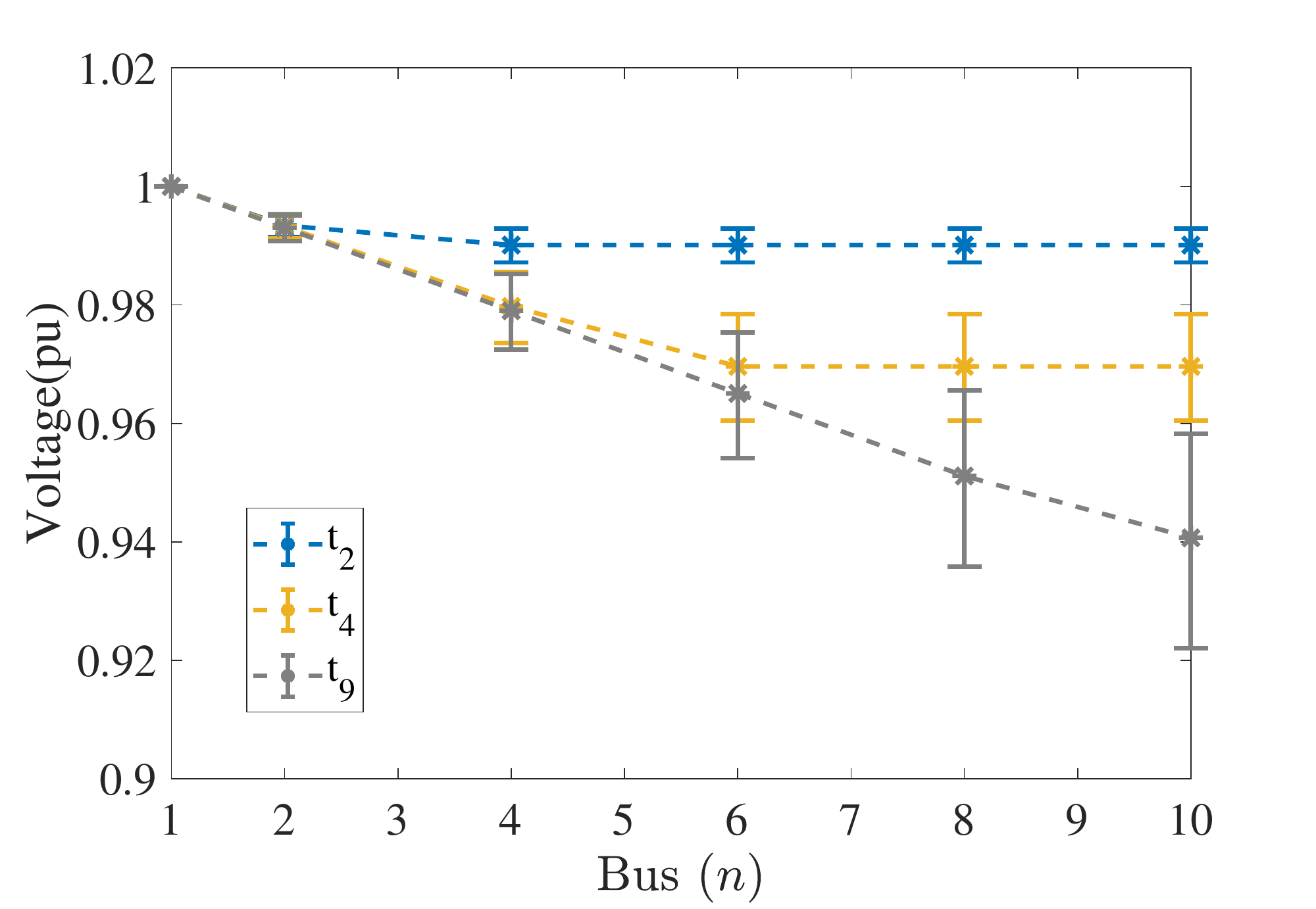}
    \caption{Voltage profile of WDC considering more general settings (one direction)}
    \label{fig:changingload}
\end{minipage}
\end{figure*}

As shown in Figure \ref{fig:voltage_one_direction}, the voltage profiles, which collect the nodal voltages at the reference bus (node $1$) to the end node (node $10$), are constructed for nine different moments, called time steps, from $t_1$ to $t_9$. This set of voltage profiles resembles half-leaf veins. Each secondary vein is a voltage profile of a time step, while the main vein is the lower bound of all profiles which will be defined later in this section. The nodal voltage reaches its minimum value at time step $t_9$ when vehicles reach the farthest nodes from the reference bus.

When the EV fleet moves towards the end of the road, the nodal voltage decrease monotonically as the ``travel'' impedance between the slack bus and the fleet's position increases so does the voltage drop. The lower bound of these voltage profiles can be characterized analytically as $V = V_1 - I_{max} Z$. Here, $V_1$ is the constant voltage of the reference node, and $I_{max}$ is the largest effective current which corresponds to the heaviest loading condition happening at the final time step as the vehicles approach the end node. $Z$ is the effective ``travel'' impedance which scales linearly as the vehicles travel and thus it is proportional to the bus number in our experiment. The lower bound is plotted as the dashed red line or the main vein of the half-leaf veins in Figure \ref{fig:voltage_one_direction}. The monotonic voltage profiles and the main vein can design voltage regulation.

Although the deployed constant-power load model is simple, it is sufficient to capture the most fundamental pattern of the voltage profile along the electrified road when the vehicles move. This simple model, however, can be extended to more complicated settings with charging efficiency, variable power consumptions, and variable speeds. For instance, we can model the effective power of a vehicle as $P=P_{EV}-\Delta{P}$ and $Q=Q_{EV}-\Delta{Q}$ where $P_{EV}$ and $Q_{EV}$ are the base power consumption. We introduce $\Delta{P}$ and $\Delta{Q}$ to represent the power variation, for example, due to non-unity charging efficiency and variable speeds. Figure \ref{fig:changingload} plots three representative voltage profiles with error bars as the power variation can vary randomly within $50\%$ of the base power level. The error bars become wider if the variation range gets larger. Moreover, for a given variation range, one still can construct the bounds of voltages, by collating the corresponding extreme voltage values.
\subsubsection{WDC EVs system with voltage compensators and PVs} 
\begin{figure}[h]
    \centering
    \includegraphics[width=1\columnwidth]{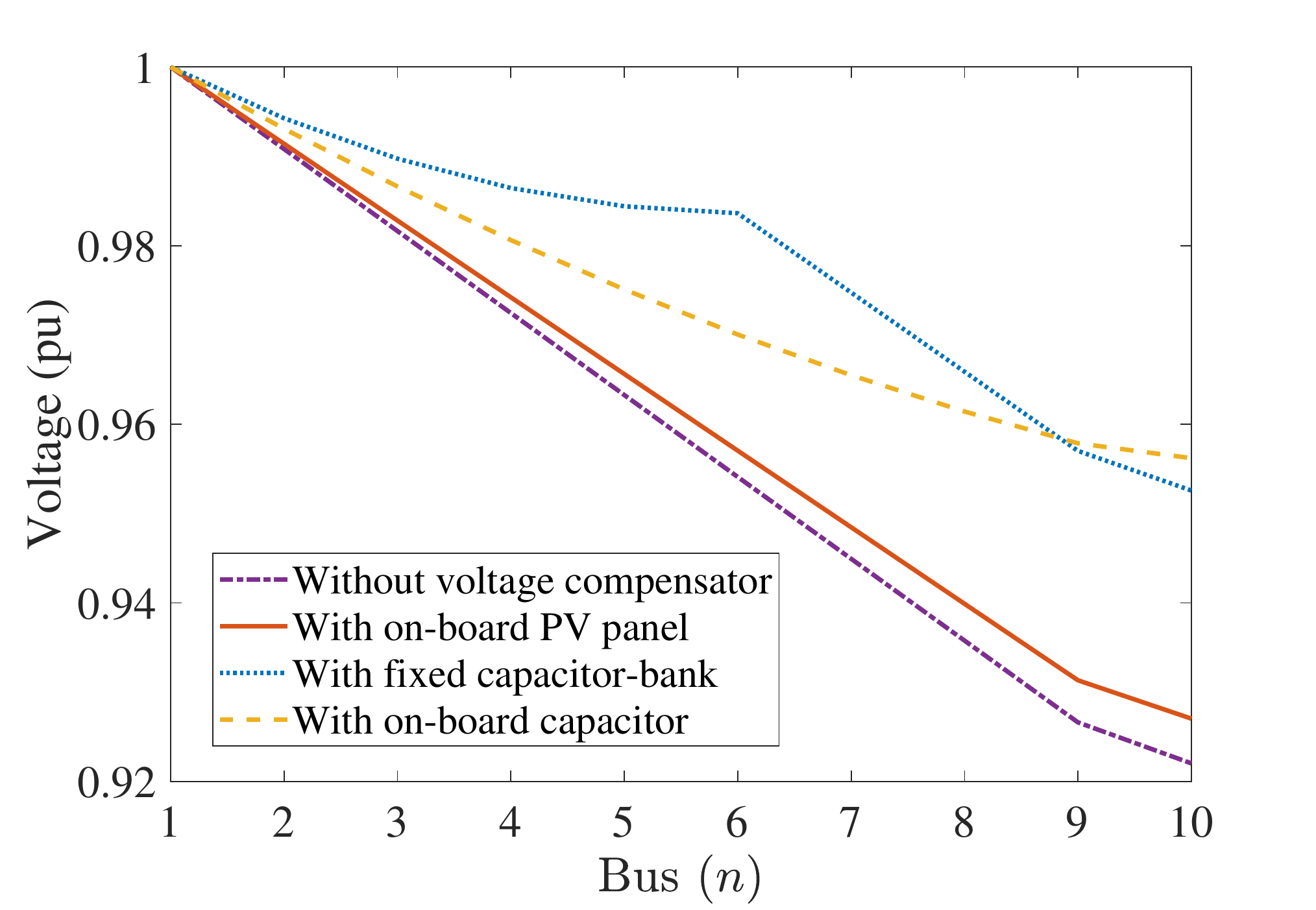}
	\caption{Voltage profile at the final time step $t_9$}
	\label{fig:compare_one_direction}
\end{figure}
In this experiment, PV panels or on-board capacitors are installed on each vehicle, whereas a fixed capacitor bank is set at bus 6--the middle of the road. For the on-board PV panel, the active power variation $\Delta P$ is used to represent the amount of PV injection. As PV produces more power, the effective power consumption of the vehicle becomes smaller. The on-board capacitor can be modeled in a similar way with the reactive power variation $\Delta Q$. However, unlike the on-board devices, the fixed capacitor bank will not change the effective power consumption of the vehicles, but the self-susceptance of the associated node/bus where the capacitor bank is located \cite{matpowermanual}.

In Figure \ref{fig:compare_one_direction}, we plot several voltage profiles to show the effect of added PVs and voltage compensators. The lowest voltage profile represents the case without PVs and voltage compensators, while the most effective compensation method is using a fixed capacitor bank. On-board capacitors affect the voltage of the bus where vehicles move over it, but the fixed capacitor bank regulates voltage locally around bus 6. 
\subsection{Two-way Road}
\begin{figure}[h]
    \centering
    \includegraphics[width=1\columnwidth]{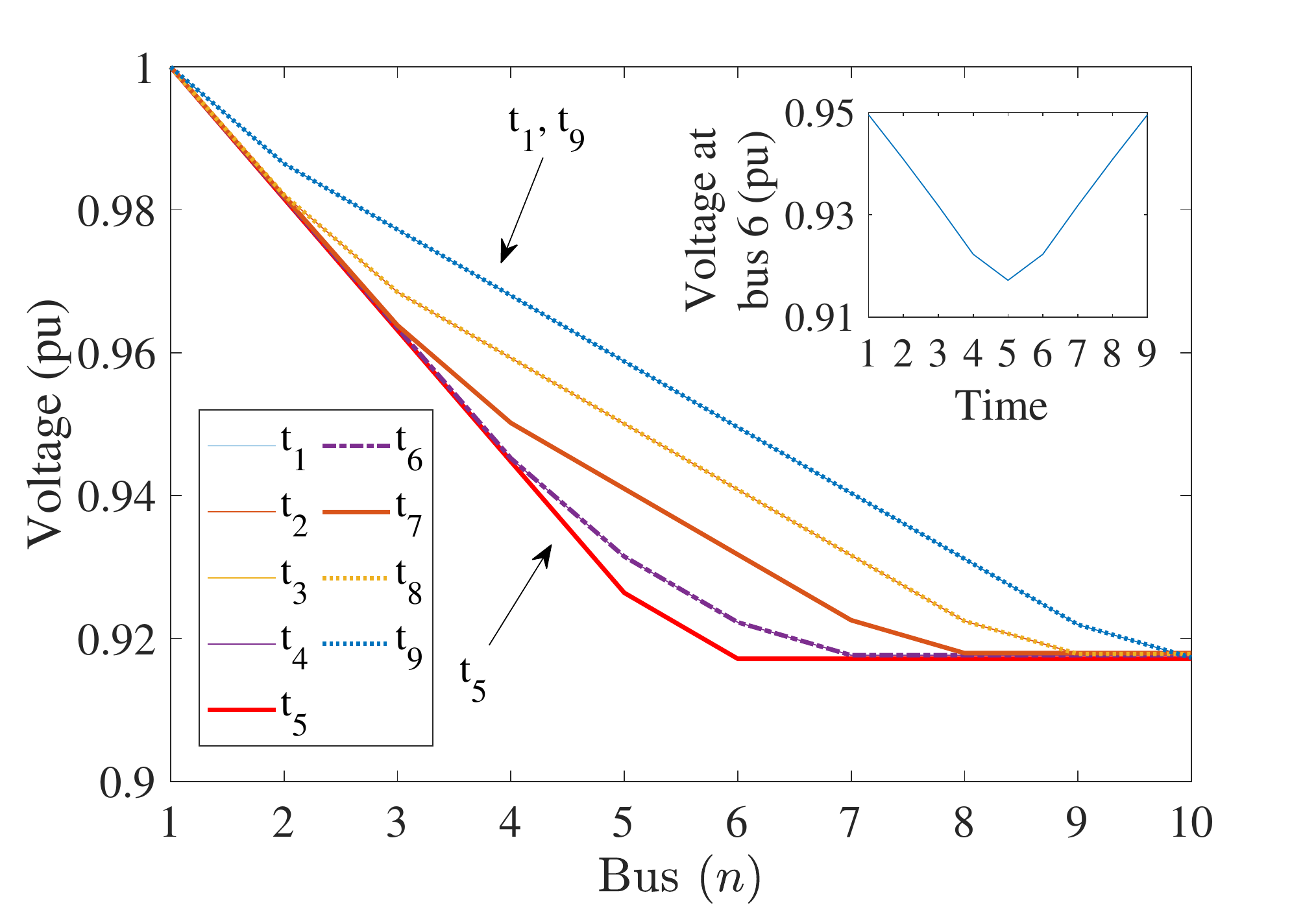}
	\caption{Voltage profile of WDC system (two directions)} 
	\label{fig:voltage_two_directions_add}
\end{figure}
In this section, a two-vehicle fleet moves from the reference bus towards the end bus, while another fleet of two vehicles moves from the opposite direction. At the beginning time, each pair of vehicles is located at the first and the last two buses, respectively.

Figure \ref{fig:voltage_two_directions_add} illustrates the voltage profiles when two fleets of EVs move in two directions. Two interesting features of voltage profiles are observed in this case. First, the set of voltage profiles has a harp-like shape as all individual voltage profiles lie inside two boundaries corresponding to two special time steps. One moment is when two fleets of vehicles are located at their beginning positions; the other moment is when they meet near the middle of the road. However, if the number of vehicles in two fleets is different, the minimum value of voltage will appear when the fleet with more vehicles reaches the end bus corresponding to their moving direction. The second feature is the repetition of voltage profiles which creates voltage swings. As two fleets of vehicles meet and continue following their original directions, some voltage profiles are overlapped. This explains why some moments such as $t_1$ and $t_9$ have the same voltage profile, or in other words, some voltage profiles are repeated. The inserted figure shows the voltage fluctuation at bus 6. Other buses also exhibit voltage swings but with different magnitudes. These voltage swings may cause problems related to power quality, voltage regulation, and power transfer losses.
\section{Long-term voltage stability}\label{sec:stability}
In this part, we present a method to analyze the long-term voltage stability of the wireless dynamic charging EV system. Only one-way road is considered in this case, but one can also extend the results to two-way road easily. We rely on continuation power flow method (CPF) which has been used widely to quantify the maximum loadability corresponding to LTVS limits \cite{ajjarapu1992continuation, Hungframe2018, mehta2016numerical}. In CPF, one increases the injections gradually and then solve the resulting series of power flow problem for the voltage levels. By collating the power injections and voltage solutions, one can construct nose curves and determine the loadability limit at the tip of the nose curve \cite{instability}. However, the CPF assumes that the positions of the injections are fixed. The moving loads in WDC problem, on the other hand, may maintain a fixed level of power consumption but change their locations during the experiment period. These moving loads, therefore, make CPF not directly applicable to an electrified road with WDC, except for a simple two-bus equivalent configuration as shown in Appendix. For larger feeder networks with more EVs, one needs to apply CPF differently as below.
\subsection{Maximum Road Length with Fixed EV Fleet}
\begin{figure}[h]
    \centering
    \includegraphics[width=1\columnwidth]{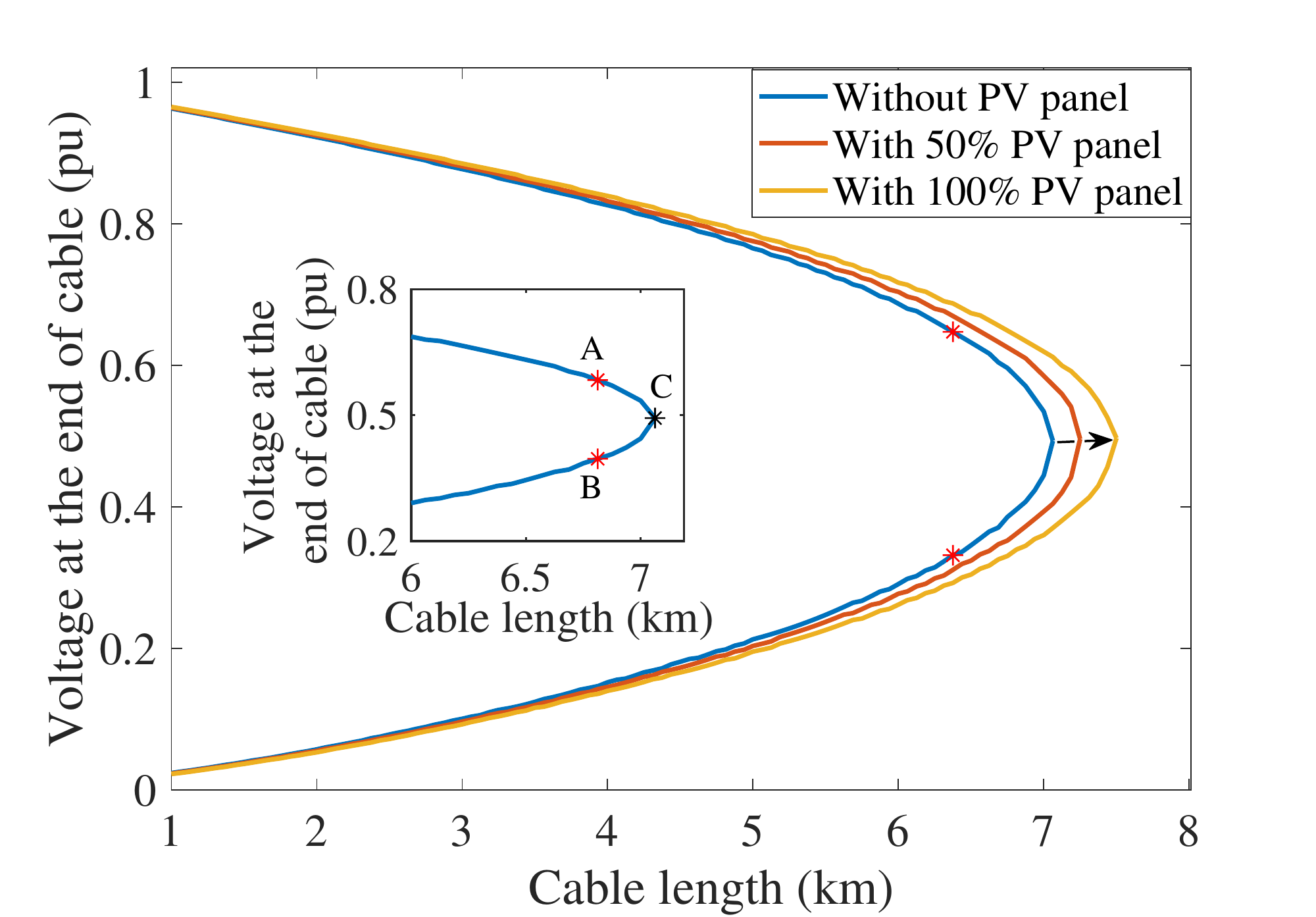}
	\caption{Maximum road length of WDC EV system. The black arrow indicates increasing PV installation.}
	\label{fig:length_add}
\end{figure}
Here, the size and power consumption of the EV fleet are fixed so that the long-term voltage stability becomes the problem of finding the maximum length of the road that corresponds to the most critical moment as vehicles reach the end of the road. The maximum length of the road can be determined, for a fixed fleet of vehicles, by gradually increasing the length of the road until no voltage solution exists. This approach is known as nose curve scenario in the typical LTVS study, but here we do not aim to construct the conventional curve of the voltage magnitude and the active power, but the voltage-road length curves. 

The detailed construction of such voltage-road length curves is as follows. First, we fix the road length and then use continuation power flow to find voltage solutions corresponding to the loading condition when all vehicles arrive at the endpoint \cite{ajjarapu1992continuation}. If the road length is less than the critical length, one can find two such voltage solutions, for example, the two red star-points in Figure \ref{fig:length_add}. Then, we gradually increase the road length and collate all resulting voltage solutions to construct the voltage-road length curve. Continue this procedure until the tip C of the voltage-road length curve where the pair of voltage solutions merge. Any road with a longer length than the critical length at point C will result in no voltage equilibrium solutions when the fleet of vehicles reach its endpoint. This situation is regarded as voltage collapse.
Figure \ref{fig:length_add} depicts a number of voltage-road length curves from which the maximum road length can be determined. The blue curve presents the base case without PV panels installed on the EVs. With more added PVs, the curves extend to the right that implies that the maximum length of the road can be larger. This effect makes sense from the steady-state study point of view. Another point to make is that the PV generation may vary depending on the weather conditions so that the maximum length can be different from the base case. This effect needs to be addressed properly in the network design and operation to ensure safe operation of the road.

\begin{figure}[h]
    \centering
    \includegraphics[width=1\columnwidth]{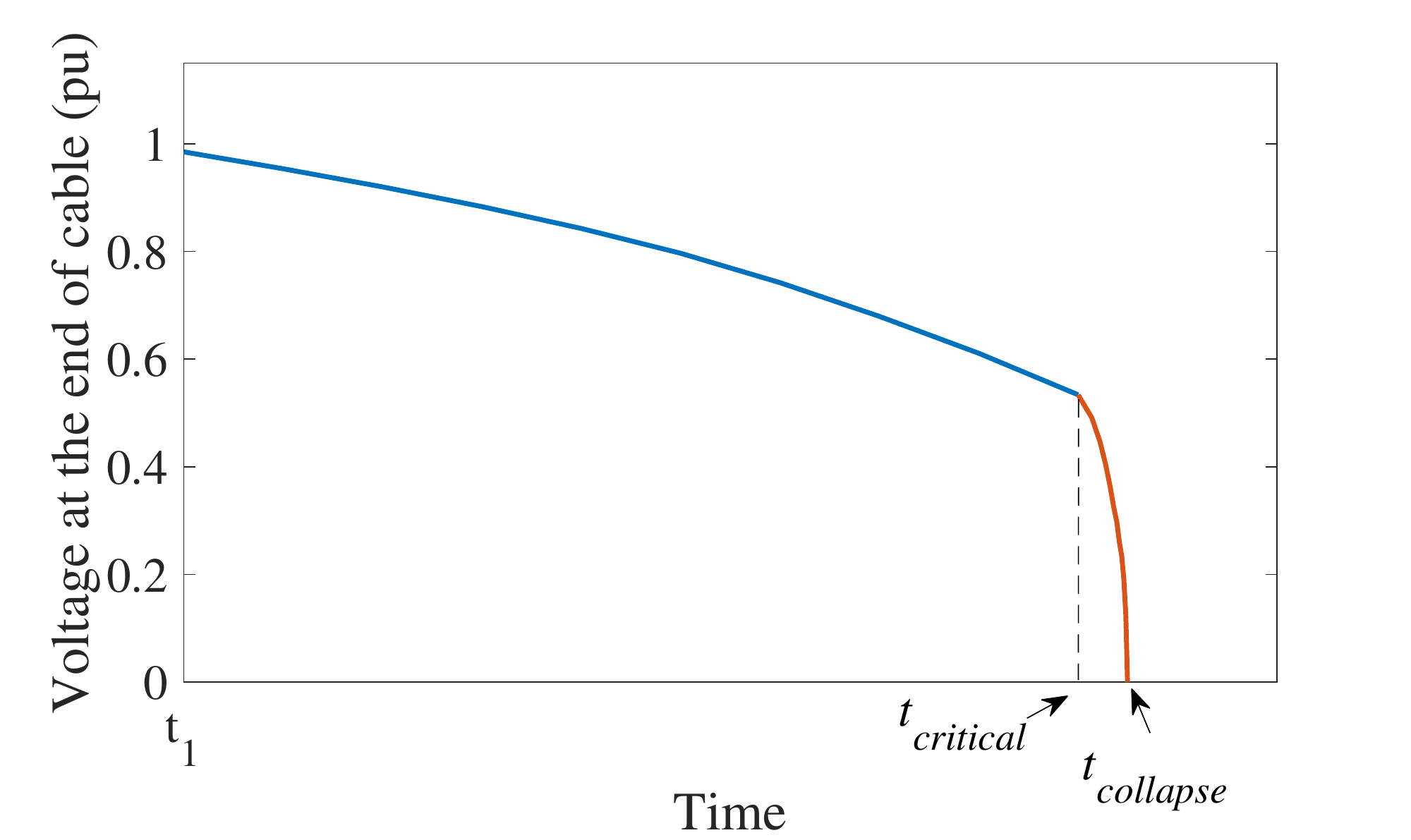}
	\caption{Voltage collapse phenomenon}
	\label{fig:collapse}
\end{figure}

While most of the experimental results are intuitive, we discuss a non-intuitive result in this section. We consider a situation when the vehicles move beyond the maximum allowed length of the road then quickly ``return back'' within the safe length. Because the vehicle has returned to the safe length, one may expect the road is ``safe''. However, Figure \ref{fig:collapse} shows an unexpected outcome: the voltage collapses and goes to zero. While the complete mechanism of this collapse behavior is rather complicated \cite{voltage}, a simple explanation is that when the vehicles pass point C towards point B in the embedded figure in Figure \ref{fig:length_add}, the solution follows the lower branch of the nose curve which contains the segment CB. Even the vehicles physically return to the safe length, the system voltage continues following the lower solution branch and does not return to point A, then finally goes to zero. This collapse phenomenon develops in a short period from the moment the vehicles arrive point C at $t_{critical}$ to $t_{collapse}$. The actual collapse event in practice has been also analysed in \cite{athen}.
\subsection{Maximum Number of Vehicles on a Fixed Length Road}
\begin{figure}[t]
    \centering
    \includegraphics[width=1\columnwidth]{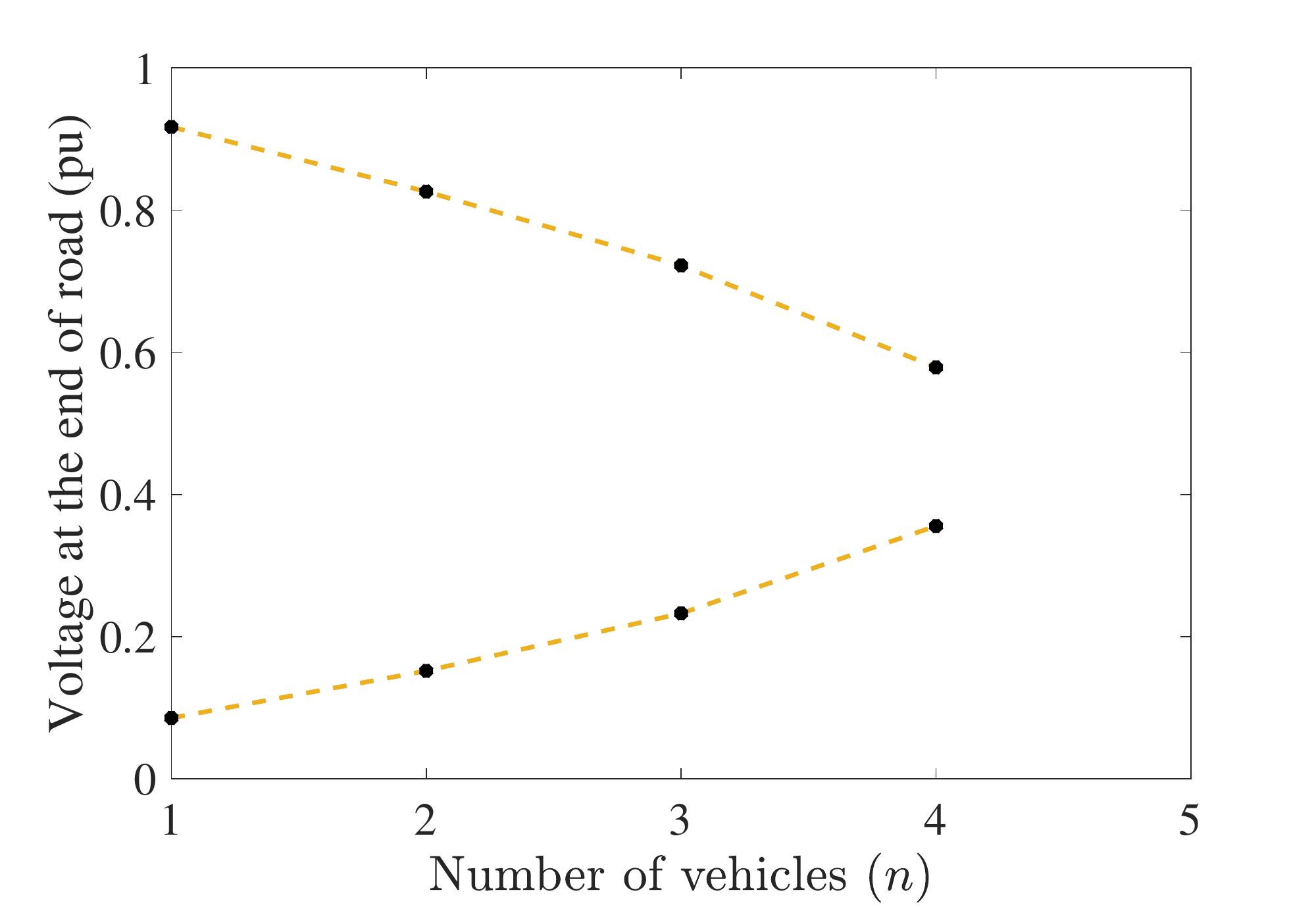}
	\caption{Maximum number of vehicles of WDC system}
	\label{fig:number}
\end{figure}
This section focuses on the maximum number of EVs allowed to operate on a fixed length road. From Figure \ref{fig:voltage_one_direction}, we observe that the heaviest loading condition is when vehicles approach the end of the road. We can use the same method for constructing the voltage-road length curves in the previous section to build the voltage-number of EVs counterparts, by gradually increasing the number of EVs operating on the road. One difference, in this case, is that the number of EVs is not continuous but has discrete values. Figure \ref{fig:number} shows that at most four EVs can be allowed to operate simultaneously on a 4-km road. Varying segment sizes can change both the maximum length of the road and the maximum number of vehicles. In our experiments, we assume that all vehicle members in a fleet follow one another and that the gap between two consecutive members is equal to the segment size. If the segment size reduces, vehicle members run closer to each other and vice versa. Reducing the segment size will make the fleet closer to the road end at the final time step and thus burden the electrified road further. Therefore, with a smaller segment size, the maximum road length and the number of vehicles allowed to operate need to be reduced.
\section{Conclusions and future work}
In this paper, we study the effect of moving loads on the steady-state operation of wireless dynamic charging system modeled as one-feeder distribution girds. Voltage profiles of various operating conditions with PV panels and voltage compensators are constructed. We also present a method to apply continuation power flow for the LTVS of electrified roads to characterize the maximum length and maximum number of EVs that the road can accommodate. The new patterns of the distribution system's steady-state equilibrium regarding both the voltage and stability limits call for new research in this new paradigm. Moreover, as this work deploys a rather simple model for the steady-state analysis of one electrified road, we plan to extend it to dynamic analysis with more mixed, detailed models of the distribution network in future.

\section{Acknowledgement}
Work of CW and HN was supported by NTU SUG. We thank Yashar Ghiassi for his shepherding.

\bibliographystyle{IEEEtran}
\bibliography{main.bib}
\section{Appendix} \label{sec:app}

\subsection{Power Flow Equation} \label{sec:apppf}
The power flow equation can be derived from Kirchhoff's circuit laws and used to describe the steady state condition of a power system. For bus $i$, the equation is as follows:
\begin{equation}\label{eq:pf}
P_i + j Q_i
= \sum_{k=1}^{n}V_i V_k (\cos\theta_{ik}+j \sin \theta_{ik})(G_{ik}-jB_{ik})
\end{equation}
where $Y_{ik}=G_{ik} + j B_{ik}$ represent the admittance of branch $ik$, $v_i = V_i \exp{(j\theta_i)}$ is the nodal complex voltage at bus $i$, and $\theta_{ik} = \theta_i-\theta_k$ denotes the angle difference between bus $i$ and bus $k$ \cite{power}.

For the steady-state analysis, we solve the set of power flow equations \eqref{eq:pf} to find the nodal voltages for given loading level $P_i$ and $Q_i$, $i = 1, \dots, n$. We then construct the voltage profile along the road by recording the solution voltages at different node locations while the vehicles are passing by.

\subsection{CPF For A Two-bus System}\label{sec:appcpf}
To show that the conventional CPF can only handle moving loads in a two-bus system, we consider a toy example with one slack and one PQ bus connecting through a line with the impedance of $R+jX$. The two nodal voltages are $V_1$ and $V_2$. This configuration corresponds to the situation one vehicle travels in a one-way road. The consumption of the vehicle $P+jQ$ is fixed, but the distance it travels increases over time; this can be captured by changing the line impedance duly.
If the power factor and $X/R$ ratio are constant, a power flow equation for this toy example can be written as follow
\begin{align}\label{equation final}
   0 \,  = \,& U^2 + \left(2PR\,(\tan{\phi}\tan{\theta}+1)-(V_1)^2\right)U \nonumber  \\&+ (PR)^2 (\sec{\phi} \sec{\theta})^2
\end{align}
where $\tan{\phi}=Q/P$, $\tan{\theta}=X/R$, $U = (V_2)^2$.
Apparently, $R$ and $P$ play the same role in \eqref{equation final} so that we can change the power injection $P$ instead of the line resistance $R$ in CPF study. Unfortunately, this is not the case when we extend to more general network configurations.
\end{document}